\documentclass[prl,preprint,groupedaddress,showpacs]{revtex4-1}

\usepackage{amsmath}
\usepackage{amssymb}
\usepackage{amsfonts}
\usepackage{mathrsfs}
\usepackage{epsfig}
\usepackage{bm}
\usepackage{multirow}
\usepackage[switch]{lineno}
\begin{document}

\title{High orbital angular momentum harmonic generation}
\author{J. Vieira$^{1}$,R.M.G.M. Trines$^2$, E.P. Alves$^{1}$, R.A. Fonseca$^{1,3}$,J.T.Mendon\c{c}a$^1$,R. Bingham$^{2}$, P.Norreys$^{2,4}$, L.O.Silva$^{1}$}
\affiliation{$^1$GoLP/Instituto de Plasmas e Fus\~{a}o Nuclear,  Instituto Superior T\'{e}cnico, Universidade de Lisboa, Lisbon, Portugal}
\affiliation{$^2$Central Laser Facility, STFC Rutherford Appleton Laboratory, Didcot, OX11 0QX, United Kingdom}
\affiliation{$^3$DCTI/ISCTE Lisbon University Institute, 1649-026 Lisbon, Portugal}
\affiliation{$^4$Department of Physics, University of Oxford, Oxford OX1 3PU, UK}
\today

\begin{abstract}
%As a fundamental property of light, the orbital angular momentum (OAM) can in principle be changed independently of other laser properties. Here 
We identify and explore a high orbital angular momentum (OAM) harmonics generation and amplification mechanism that manipulates the OAM independently of any other laser property, by preserving the initial laser wavelength, through stimulated Raman backscattering in a plasma. The high OAM harmonics spectra can extend at least up to the limiting value imposed by the paraxial approximation. We show with theory and particle-in-cell simulations that the orders of the OAM harmonics can be tuned according to a selection rule that depends on the initial OAM of the interacting waves. We illustrate the high OAM harmonics generation in a plasma using several examples including the generation of prime OAM harmonics. The process can also be realised in any nonlinear optical Kerr media supporting three-wave interactions.
\end{abstract}

\pacs{52.38.Bv, 52.65.Rr, 42.65.Ky}

\maketitle

Unlike cylindrically symmetrical wavefronts, which have an intensity maximum on-axis and nearly flat wavefronts, OAM lasers have doughnut intensity profiles and helical wavefronts. Since the seminal paper by L. Allen~\emph{et al.}\cite{bib:allen_pra_1992}, these unique properties have led to many scientific and technological advances. Experiments demonstrated the generation of entangled photons with OAM, opening new directions in quantum computing~\cite{bib:mair_nature_2001}. The OAM also promises to greatly enhance optical communications~\cite{bib:wang_nphoton_2013}. It plays a pivotal role in super-resolution microscopy~\cite{sted}, allows for new kinds of optical tweezers~\cite{bib:padgett_pras_2014}, and it might be even used as a diagnostic for rotating black holes~\cite{bib:tamburini_nphys_2011}. From a fundamental perspective, all these important contributions have only been possible because the OAM is a fundamental property of light, which can be controlled as an independent degree of freedom.

There are optical processes, such as high harmonic generation~\cite{bib:franken_prl_1961} (HHG), where independent OAM and frequency manipulations are not allowed. Because energy and momentum are conserved, it is natural to assume that the nth harmonic of a laser pulse with initial OAM level $\ell$ and photon energy $\hbar \omega$ has OAM $n \ell$ and energy $n \hbar \omega$. Much efforts have been dedicated to demonstrate this hypothesis. Apart from an exceptional result that seemed to break this assumption~\cite{bib:zurch_natphys_2012}, experiments~\cite{bib:courtial_pra_1997,bib:gariepy_prl_2014,bib:arxiv2016} and theoretical modelling~\cite{bib:mendonca_pop_2015,bib:gordon_ol_2009,bib:sobhani_jpd_2016} confirmed energy and momentum conservation in HHG, which then became the dominant view of HHG using vortex lasers.

In contrast, in this Letter we identify a Raman scattering process to create high OAM harmonics that preserves the laser frequency and total angular momentum, thereby manipulating the OAM independently of any other laser property. An interesting path to produce lasers with very high OAM levels, in which the OAM and frequency harmonics are still coupled, has been recently suggested~\cite{bib:gariepy_prl_2014}. As we will show, however, only by independently controlling the OAM and laser frequency can the potential of OAM for several applications (e.g. super-resolution microscopy) become fully realised.

\begin{figure}
\centering\includegraphics[width=.9 \columnwidth]{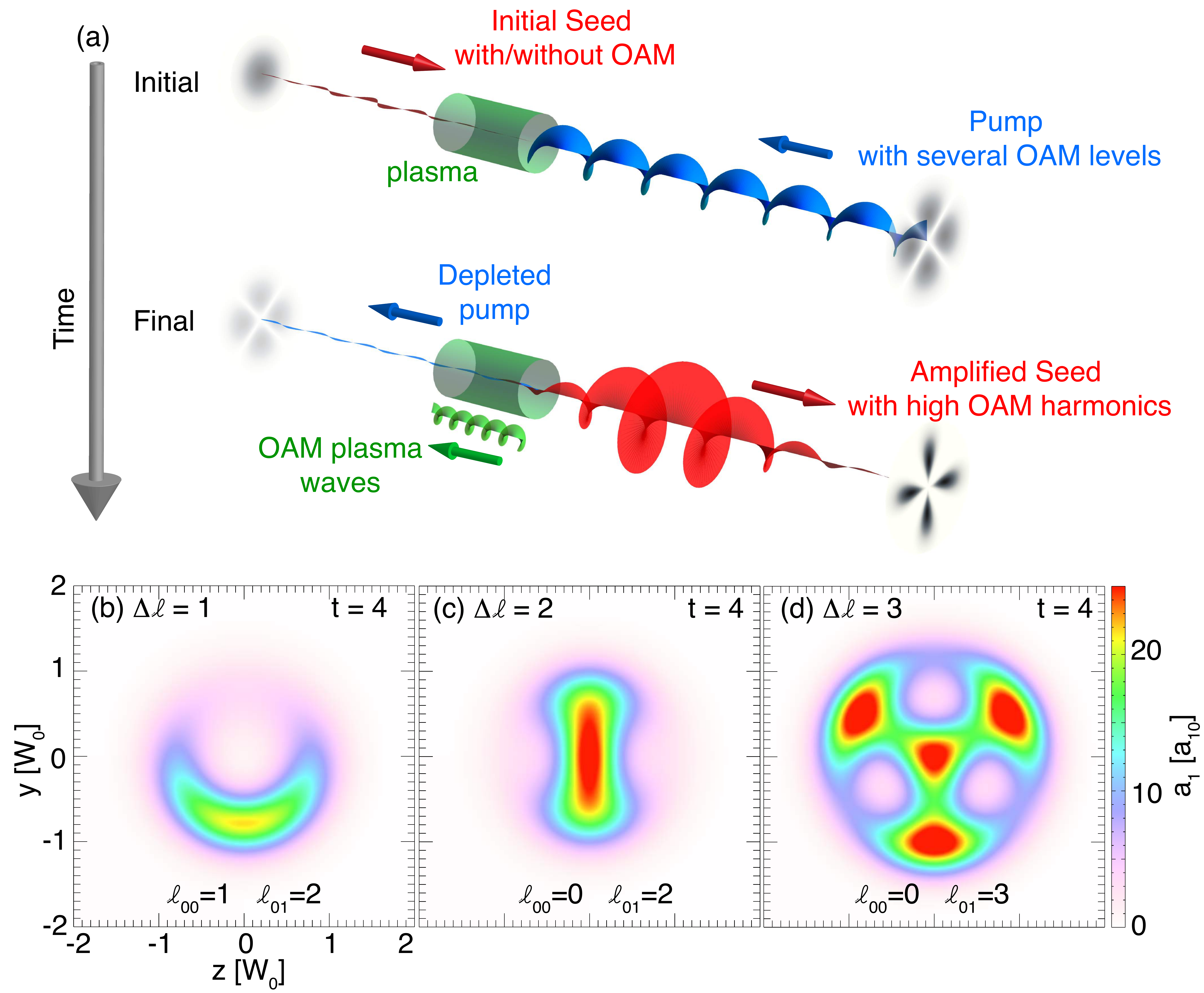}
\caption{(a) Illustration of the high OAM harmonics generation mechanism in stimulated Raman scattering. The high OAM harmonics are formed in a seed beam (red) after the interaction with a pump beam with several OAM modes in the nonlinear medium (plasma). (b)-(d) Transverse field profile of a initially Gaussian laser pulse seed after interacting with a counter-propagating pump. Initially, $a_{00}=a_{01}=a_{1}=1.0$ and time $t$ is normalised to $t = \sqrt{\alpha \omega_p/\omega_1}$. (b)-(d) show the seed pulse after interacting with a pump with $\Delta \ell = 1$ (b), $\Delta \ell = 2$ (c), and $\Delta \ell = 3$ (d) respectively.}
\label{fig:figure1}
\end{figure}

We consider a three-wave interaction mechanism, stimulated Raman scattering, in a plasma~\cite{bib:forslund_pf_1975,bib:shvets_prl_1998,bib:malkin_prl_1999,bib:ren_nphys_2007,bib:trines_nphys_2011,bib:trines_prl_2011,bib:medonca_prl_2009,bib:fraiman_pop_2002}. Raman scattering has been investigated in the frame of one-dimensional physical models~\cite{bib:clark_pop_2003,bib:berger_pop_2004}, focusing on energy flows between the intervening waves. As an application, we recently showed that Raman backscattering can amplify seed pulses with a single OAM level to the required intensities to explore relativistic laser-plasma interactions~\cite{bib:mendonca_pop_2014,bib:vieira_prl_2014,bib:brabetz_pop_2015,bib:mendonca_prl_2009}, by using a counter-propagating long pump laser that also contains a single OAM mode~\cite{bib:vieira_natcomms_2016}. Here we counter this view, by demonstrating that three-wave interactions can also be used to manipulate, with an unprecedented degree of controllability, the three-dimensional spatial-temporal laser pulse structure. Specifically, we show that the seed will gain high OAM harmonics whilst preserving its carrier frequency if the pump contains several modes with different OAM levels (Fig.~\ref{fig:figure1}a). The OAM harmonics extend, at least, up to the paraxial limit. These results demonstrate that the scope of stimulated Raman scattering and, more generally, three-wave processes, goes beyond laser amplification by enabling an unexplored type of spatial-temporal control that has never been considered. Because the technique controls the OAM as an independent degree of freedom, a combination of the conventional HHG scheme and the Raman scheme opens the way for producing lasers with extremely high OAM levels, and, simultaneously, very high photon frequencies.

The order of the high OAM harmonics follows a simple algebraic expression only involving the initial OAM of the intervening lasers, being given by $\ell=\ell_{1}+m \Delta \ell $, where $\ell_1$ is the initial seed OAM, $\Delta \ell = \ell_{00}-\ell_{01}$ is the OAM difference between two pump modes. The high OAM harmonics result from the angular momentum cascading from the modes with lower OAM to the modes with higher OAM. This is an unexpected and unexplored behaviour, which is profoundly related to the role of the plasma wave as a spiral phase element. Thus, similar mechanisms are expected when dealing with waves or oscillations in other non-linear media, e.g. phonons or molecular vibrations. We confirm the analytical predictions numerically through \emph{ab-initio} three-dimensional OSIRIS simulations~\cite{bib:fonseca_book,bib:osiris}. OSIRIS employs the particle-in-cell (PIC) technique, taking into account the kinetic plasma response at the single particle level, in the presence of the external and self-consistently generated electromagnetic fields and without any physical approximations to the extent where quantum effects can be neglected.

In the small signal regime, valid when the amplitude of the seed is much smaller than that of the pump, the evolution of the seed envelope $a_1$ is given by~\cite{bib:vieira_natcomms_2016}:
\begin{equation}
\label{eq:seed}
\mathbf{a}_{1}(t) = \left(\mathbf{a}_{1}(t=0)\cdot \frac{\mathbf{a}_{0}^*}{|\mathbf{a}_{0}|}\right) \frac{\mathbf{a}_{0}}{|\mathbf{a}_{0}|} \cosh\left(\Gamma t\right),
\end{equation}
\begin{equation}
\label{eq:gamma}
\Gamma^2  = \frac{e^2 k_p^2 \omega_p^2}{8 \omega_0 \omega_{1}m_e^2} |\mathbf{a}_{0}|^2,
\end{equation}
where $a_0$ is the envelope of the pump, $\omega_p$, $\omega_0$ and $\omega_1$ are the frequencies of the plasma wave, pump and seed lasers, respectively (corresponding wavelengths are given by $\lambda_p$, $\lambda_0$ and $\lambda_1$ respectively). In addition, $e$ and $m_e$ are the electron charge and mass, $k_p = \omega_0 (2 - \omega_p/\omega_0 - \omega_p^2/\omega_0^2) \sim 2 \omega_0 - \omega_p$ is the plasma wavenumber, and $c$ the speed of light. Equations~(\ref{eq:seed}) and (\ref{eq:gamma}) are similar to the solutions of other three-wave interaction processes in the small signal regime, such as frequency sum/difference generation for instance.
%, with the only difference being the multiplicative factors. 
Equations~(\ref{eq:seed}) and (\ref{eq:gamma}) also describe stimulated Raman backscattering of lasers with arbitrary polarisations (although seed and pump need to have non-orthogonal polarisation components for Raman to occur) and arbitrary transverse field envelopes profiles. %These equations predict the generation and amplification of single OAM level lasers~\cite{bib:vieira_natcomms_2016} to the powers required to explore relativistic laser-plasma interactions~\cite{bib:mendonca_pop_2014,bib:vieira_prl_2014,bib:brabetz_pop_2015,bib:mendonca_prl_2009}.
Here we focus on the scenario where the lasers are linearly polarised in the same direction, and where the pump laser contains several Laguerre Gaussian modes.

Unlike lasers described by a single OAM level, the time-averaged intensity envelope of a laser containing several OAM levels is no longer cylindrically symmetric. Consider, for instance, the superposition between two modes with $\ell_{00}$ and $\ell_{01}$ each with peak vector potential given by $a_{00}$ and $a_{01}$. The corresponding transverse intensity envelope ($I_0(\mathbf{r})\propto a_0 a_0^*)$, given by $I \propto a_{00}^2 + a_{01}^2 + 2 a_{00} a_{01} \cos\left[\left(\ell_{00}-\ell_{01}\right)\phi\right]$, contains a cylindrically symmetric term ($a_{00}^2 + a_{01}^2$), corresponding to the sum of the intensity envelopes of each individual Laguerre-Gaussian component, and a non-cylindrically symmetric component ($2 a_{00} a_{01} \cos\left[\left(\ell_{00}-\ell_{01}\right)\phi\right])$ that corresponds to the beating of the two Laguerre-Gaussian modes. This analysis shows that non-cylindrically symmetric pump lasers resulting from the combination of several OAM modes change key properties of stimulated Raman backscattering because the growth rate, given by Eq.~(\ref{eq:gamma}), now depends on $\phi$. The $\phi$ dependent growth rate can significantly re-shape the intensity of the seed, as shown in Fig.~\ref{fig:figure1}b-d. Figure~\ref{fig:figure1}b-d shows the final intensity profile of an initially Gaussian seed pulse ($\ell_1 = 0$) using a pump that is a superposition of OAM states according to Eqs.~(\ref{eq:seed}) and (\ref{eq:gamma}). The seed transverse envelope develops evenly spaced lobes at well defined locations along the azimutal $\phi$ direction. The number of lobes corresponds to $\Delta \ell \equiv \ell_{00}-\ell_{01}$. For $\Delta \ell = 1$, a single lobe appears (Fig.~\ref{fig:figure1}b). When $\Delta \ell = 2$, two lobules are formed spaced by $\Delta \phi = \pi$ (Fig.~\ref{fig:figure1}c), and when $\Delta \ell = 3$ there are three lobules separated by $\Delta \phi = 2 \pi / 3$ (Fig.~\ref{fig:figure1}d). We note that $a_{00}$ and $a_{01}$ (and $a_1$) are the local, radius dependent amplitudes of each mode.

The laser envelopes shown in Fig.~\ref{fig:figure1}b-d contain high OAM harmonics. High OAM harmonics generation in stimulated Raman scattering results directly from the conservation of OAM. Initially, the seed is described by a single Laguerre-Gaussian mode with $\ell_1$, and the pump described by two OAM modes $\ell_{00}$ and $\ell_{01}$. This seed mode beats with each pump mode exciting a Langmuir plasma wave. In order to conserve angular momentum, the two modes in the plasma have $\ell_{00}-\ell_1$ and $\ell_{01}-\ell_1$, i.e. the modes in the plasma absorb the OAM difference between every pump mode and the seed. The beating of these two plasma modes with the pump is at the onset of the cascading mechanism that leads to the high OAM harmonic generation. The pump $\ell_{00}$ beating with the plasma $\ell_{01}-\ell_1$ preserves the initial OAM if a new seed component appears with $\ell_1 + \Delta \ell$ ($\Delta \ell \equiv \ell_{00}-\ell_{01}$). Similarly, the pump $\ell_{01}$ beating with the plasma $\ell_{00}-\ell_1$ preserves the OAM if another seed mode component grows with $\ell_1 - \Delta \ell$. These are the first OAM sideband harmonics growing in the seed. Each of these sidebands will beat with the pump, adding new higher-order OAM sidebands to the  plasma wave. In general, each new mode in the plasma wave continues to interact with every pump mode, generating higher OAM seed modes $\ell_1 \pm m \Delta \ell$, with integer $m$. The high OAM harmonics then appear because the plasma wave is in a superposition of OAM states. This unexpected behaviour, due to the wave-like plasma response, has no counter-part in other optical devices such as spiral wave plates.

An analytical theory for the evolution of each seed OAM harmonic supports this qualitative interpretation. Instead of taking an azimuthal Fourier transform of Eq.~(\ref{eq:seed}), which cannot be determined analytically, we start with the general differential equation that describes the evolution of the linearly polarised seed in the small signal regime, $\mathrm{d}_t^2 a_1 = \alpha a_1 (a_0 a_0^*) = \alpha a_1 \left[a_{00}^2+a_{01}^2 + 2 a_{00} a_{01} \cos\left(\Delta \ell \phi \right)\right]$, where $\mathrm{d}_t^2 = \mathrm{d}^2/\mathrm{d} t^2$ and $\alpha = (e^2 \omega_p^3 /8 m_e^2 c^2 \omega_1) = \Gamma / |\mathbf{a_0}|$. Instead of focusing in the amplification properties of stimulated Raman scattering, we describe, theoretically, the evolution inner three-dimensional spatial-temporal OAM structure of the seed. The evolution of the amplitude of each OAM component, $f_m = (1/2\pi)\int_0^{2\pi} a_1 \exp\left(-i m \phi \right)\mathrm{d}\phi$, can be readily obtained by multiplying the equation by $(1/2\pi)\exp\left(-i m \phi \right)$ and then by integrating over $\phi$. This procedure leads to the following differential equation for the evolution of each seed mode:
\begin{equation}
\label{eq:oamharmonics}
\frac{\mathrm{d}^2 f_m}{\mathrm{d} t^2} = \frac{e^2 \omega_p^3 }{8 m_e^2 c^2 \omega_1} \left[\left(a_{00}^2+a_{01}^2\right) f_m + a_{00} a_{01} \left(f_{m+\Delta \ell}+f_{m-\Delta \ell}\right)\right].
\end{equation}

Equation~(\ref{eq:oamharmonics}) predicts the generation of high OAM harmonics. Unlike in conventional high (frequency) harmonic generation, however, the frequency of the laser remains unchanged in Eq.~(\ref{eq:oamharmonics}). %since the process creates purely spatial OAM harmonics.
According to Eq.~(\ref{eq:oamharmonics}), each mode $m$ grows from of its (unstable) coupling with the plasma and the pump [first term on the left-hand-side of Eq.~(\ref{eq:oamharmonics})]. In addition, each mode is also driven by its closest sideband mode with $\ell = m\pm \Delta \ell$ [second term on the left-hand-side of Eq.~(\ref{eq:oamharmonics})]. We can clarify the predictions of Eq.~(\ref{eq:oamharmonics}) by distinguishing three cases: At early times the fundamental mode $\ell_1$ is driven by the first term on the right-hand-side (rhs) of Eq.~(\ref{eq:oamharmonics}) even if the second term is not immediately available for these modes. Then, sideband modes at $\ell_1 \pm |m|\Delta \ell$ are created and amplified once the second term on the rhs of Eq.~(\ref{eq:oamharmonics}) appears because the first term is not initially available. The modes between $\ell_1 \pm |m|\Delta \ell$ and $\ell_1 \pm |m \pm 1|\Delta \ell$ will never grow since neither the first or the second term on the rhs of Eq.~(\ref{eq:oamharmonics}) will ever be available.

\begin{figure}
\centering\includegraphics[width=\columnwidth]{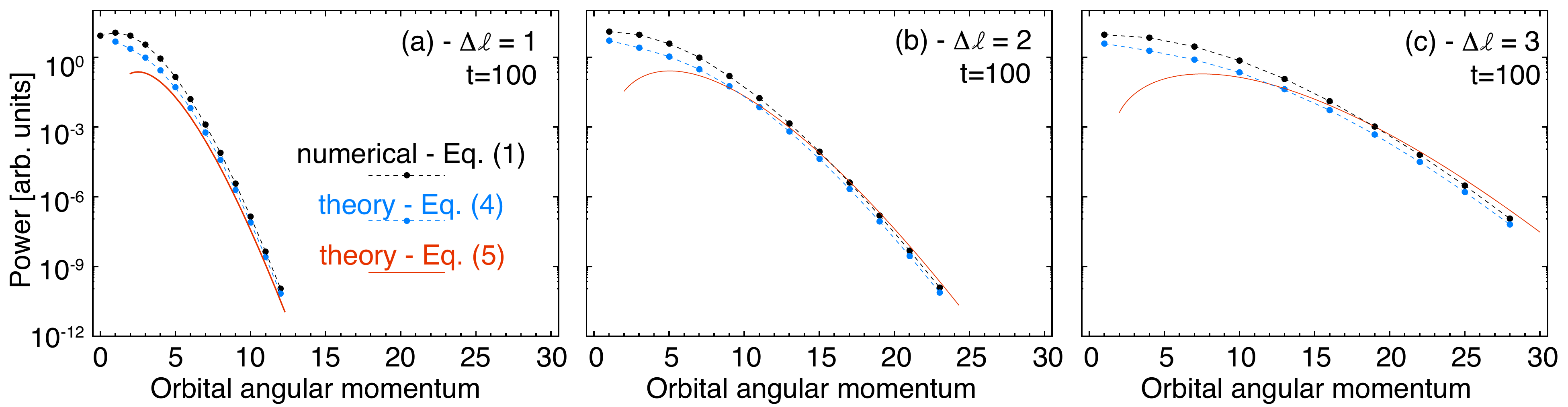}
\caption{Comparison between the numerical [black - Eq.~(\ref{eq:seed})] and theoretical prediction [blue - Eq.~(\ref{eq:oamsolution}) and red - Eq.~(\ref{eq:oamsolution})] for the generation of the high OAM harmonics Fourier power spectra. Calculations assumed $a_{00}=a_{01}=0.1$ and $a_1 = 0.01$, and normalised growth rate $\Gamma = |\mathbf{a_0}|$, and time is normalised to $\sqrt{\alpha \omega_p/\omega_1}$. Initially the seed pulse has $\ell_1 = 1$. The pump has $\Delta \ell = 1$ in (a); $\Delta \ell = 2$ in (b) and  $\Delta \ell = 3$ in (c).}
\label{fig:figure2}
\end{figure}

The evolution of the high OAM harmonics at sufficiently early times, for which the higher OAM sidebands are driven by their closest neighbour OAM mode, is:
\begin{equation}
\label{eq:oamsolution}
a_1^{\ell_1 \pm |m| \Delta \ell }(t) = a_1\left(\frac{a_{00} a_{01}}{a_{00}^2+a_{01}^2}\right)^{|m|} \left( \cosh\left(\Gamma t\right) - \sum_{i=0}^{|m|-1}\frac{(\Gamma t)^{2 i}}{(2i)!}\right),
\end{equation}
where $\Gamma^2 = e^2 k_p^2 \omega_p^2/(8\omega_0 \omega_1 m_e^2) \left(a_{00}^2 + a_{01}^2\right)$. Figure~\ref{fig:figure2} compares the numerical solution to Eq.~(\ref{eq:seed}), valid in the small signal approximation, with the analytical solution of Eq.~(\ref{eq:oamharmonics}) given by Eq.~(\ref{eq:oamsolution}), which considers that the growth of each mode is only driven by the amplitude of the preceding sideband. The comparison assumed spatially uniform $a_{00}$, $a_{01}$, and $a_{1}$. The agreement is excellent at early times and holds for different pump OAM compositions $\Delta \ell$ as long as the growth of the OAM harmonics are driven by their closest neighbours. A subsequent approximation to Eq.~(\ref{eq:oamsolution}) can be obtained by retaining the leading order term of the factor between the big brackets in Eq.~(\ref{eq:oamsolution}):
\begin{equation}
\label{eq:a1}
a_1^{\ell \pm |m|\Delta \ell} \simeq a_1 \left[\frac{a_{00} a_{01}}{a_{00}^2+a_{01}^2}\right]^{|m|} \frac{(\Gamma t)^{2|m|}}{(2 m)!}
\end{equation}
Figure~\ref{fig:figure2} compares the numerical solution of Eq.~(\ref{eq:seed}) and the analytical solution given by Eq.~(\ref{eq:a1}). The agreement is very good for $m\gtrsim 4$. 

Equations~(\ref{eq:oamharmonics}) and (\ref{eq:oamsolution}) predict OAM harmonic generation for arbitrarily high OAM orders. Our theory, however, is not valid beyond the paraxial approximation, which can be employed as long as the longitudinal laser wavenumber, $k_{\|}$, is much higher than the transverse laser wavenumber. This assumption no longer holds when the azimutal wavenumber $k_{\phi} \sim \ell/r_0$ becomes of the same order of $k_{\|}$, i.e. when $|k_{\phi}| \simeq k_{\|}$ ($r_0 = w_0 \sqrt{\ell}$ is the distance from the origin to the radius of maximum intensity, $w_0$ is the laser spot-size). Since $k_{\|} = 2 \pi /\lambda_0$, the high OAM harmonics generation will differ from the predictions of Eq.~(\ref{eq:seed}) when $|\ell| \simeq r_0 \omega_1/c \simeq 2 \pi r_0/\lambda_1$.

\begin{figure}
\centering\includegraphics[width=.9 \columnwidth]{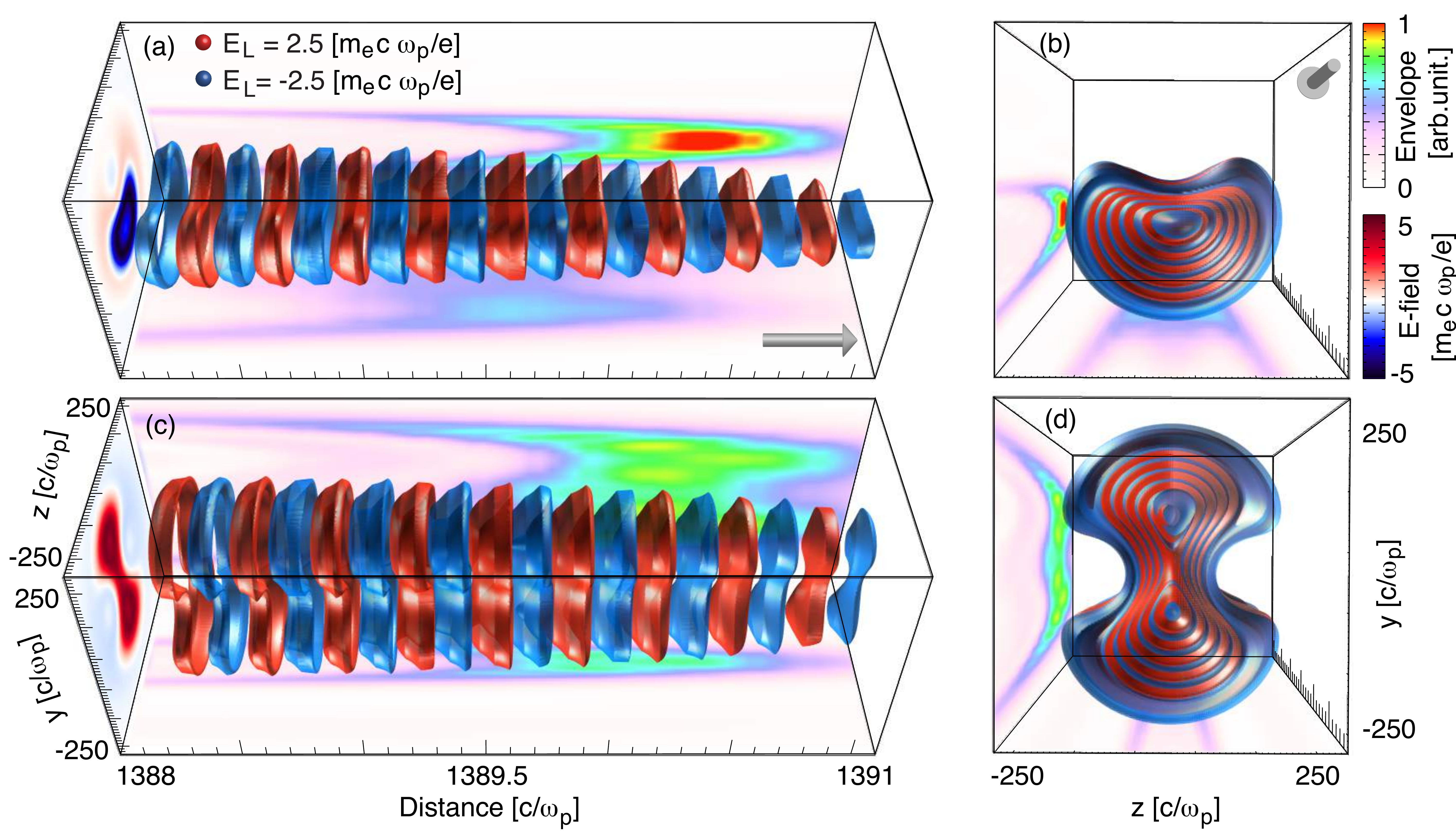}
\caption{PIC simulation results showing the seed pulse electric field containing high OAM harmonics while it is still in the plasma (not shown). The seed pulse is Gaussian. The pump has $\Delta \ell = 1$ in (a)-(b) and $\Delta \ell = 2$ in (c)-(d). Seed electric field ($E$) isosurfaces of the seed pulse shown in blue ($E<0$) and in red ($E>0$). The x-y and x-z projections show the central envelope of the seed electric field. The y-z projection shows the laser electric field at a longitudinal position $(x)$ close to the laser peak intensity. The arrows indicate the laser propagation direction. Distances are in units of $1~c/\omega_p = 2.5~\mu m$.}
\label{fig:figure3}
\end{figure}

The high OAM harmonic orders are produced according to the simple algebraic relation $\ell_1 + m \Delta \ell$. All OAM harmonics will be then generated when $\Delta \ell = 1$. Figure~\ref{fig:figure1}a and Fig.~\ref{fig:figure2}a
illustrate this scenario. Even harmonics appear when $\Delta \ell = 2$ with even $\ell_1$ (Fig.~\ref{fig:figure2}b). When $\ell_1$ is odd, then odd OAM harmonics appear instead. To illustrate high OAM harmonics generation, we have also performed three-dimensional OSIRIS PIC simulations~\cite{bib:fonseca_book,bib:osiris}. The waists of the intervening OAM modes were chosen to ensure significant overlap at the start of the interaction. Since the interaction length is much shorter than the Rayleigh lengths, there is significant overlap during propagation. Figure~\ref{fig:figure3} shows the seed electric field using $\Delta \ell = 1$ [Fig.~\ref{fig:figure3}a-b] and $\Delta \ell = 2$ [Fig.~\ref{fig:figure3}c-d] in the pump. The initial OAM of the Gaussian seed ($\ell_1 = 0$) does not change because the amplitude of each pair of positive and negative sideband harmonic is the same. Thus, the initially Gaussian ($\ell_1 = 0$) seed pulse phase fronts remain flat during the process, as shown in Fig.~\ref{fig:figure3}a and in Fig.~\ref{fig:figure3}c. For $\ell_1 \ne 0$, the phase fronts would be twisted. Each positive and negative high OAM harmonic creates an azimutal beat-wave pattern that modulates the initial seed pulse transverse field profile. These transverse modulations, shown in Figs.~\ref{fig:figure3}b,d confirm the growth of positive and negative OAM harmonics. They are also very similar to the theoretical predictions shown in Figs.~\ref{fig:figure1}a-b, supporting the validity of the theoretical model.

\begin{figure}
\centering\includegraphics[width=.9 \columnwidth]{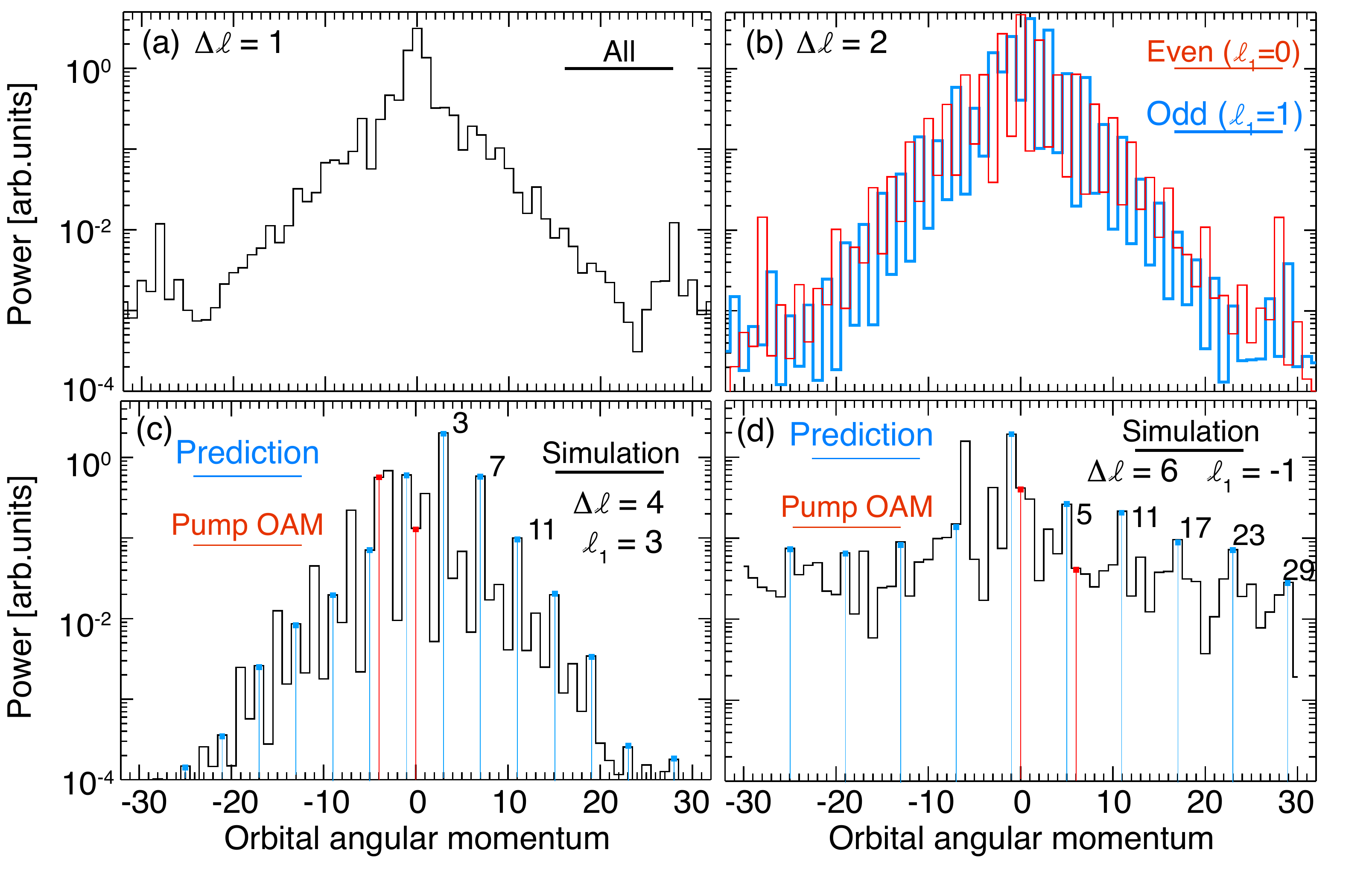}
\caption{PIC simulation results showing high OAM harmonics spectra at $z=1386.71 c/\omega_p$ (a), $z=1389.01 c/\omega_p$ (b), $z=1450.14~c/\omega_p$ (c) and $z=1443.16~c/\omega_p$ (d).}
\label{fig:figure4}
\end{figure}

Figure~\ref{fig:figure4} shows the Fourier coefficients from the PIC simulations using several seed and pump OAM combinations. Figure~\ref{fig:figure4}a and the red line in Fig.~\ref{fig:figure4}b, which show the Fourier coefficients of the laser in Fig.~\ref{fig:figure3}, confirm the generation of all OAM harmonics (~\ref{fig:figure4}a) and even modes only (~\ref{fig:figure4}b, pink line), in agreement with theory. Figure~\ref{fig:figure4}b (blue line) also shows the generation of the odd OAM harmonics using $\ell_1 = 1$.

% (see Supplementary material for the details on the numerical analysis procedure)

To further demonstrate the high controllability of the process, we provide an all-optical demonstration of the Green-Tao prime number theorem~\cite{bib:primes}. The Green-Tao theorem states that for every natural number $k$, there exist natural numbers $a(k) + b(k)$ such that the numbers in the sequence $a(k), a(k) + b(k), a(k) + 2 b(k), \ldots , a(k) + (k-1) b(k)$ are all prime. This arithmetic progression corresponds to the OAM levels produced during high OAM harmonic generation with $a = \ell_1$ and $b = \Delta \ell$. Figure~\ref{fig:figure4}c shows the generation of prime numbers with $\Delta \ell = 4$ and $\ell_1 = 3$, providing (positive) primes up to 11. Theoretical predictions are very accurate for the positive OAM sideband generation. The numerical diagnostic used to decode the seed OAM harmonics does not fully discern between positive and negative OAMs. As a result, Fig.~\ref{fig:figure4}c shows the reflection of the positive OAM sidebands in the negative OAM spectral region. Figure~\ref{fig:figure4}d shows a more demanding scenario, generating primes smaller than 29, using $\Delta \ell = 6$ and $\ell_1 = -1$. In fact, Fig.~\ref{fig:figure4} shows OAM levels all prime except for $-1$ and $-25$ (if we take the modulus), giving a total of 8 OAM prime harmonics. We note that the laser spot-sizes were adjusted to maximize the overlap between the seed and pump.

We conclude by outlining a scheme that could boost current super-resolution microscopy techniques~\cite{sted}, while also highlighting the importance of controlling the OAM as an independent degree of freedom. Consider a petal laser pulse consisting of two Laguerre-Gaussian modes with OAM levels $+\ell$ and $-\ell$. The distance between consecutive intensity maxima is $d = 2 \pi r_0/\ell$, where the radius $r_0$ where the intensity is maximum is $r_0 = w_0 \sqrt{\ell}$. The distance $d$ is also a figure for the resolution of an imaging system based on these beams. Simple arithmetic calculations show that $d$ decreases to $d_{\mathrm{HHG}} = 2 \pi w_0 / \sqrt{n \ell}$ for the nth harmonic in conventional HHG. Thus, resolution increases by $\sqrt{n}$. Combining conventional HHG with the Raman scheme can reduce $d$ even more, to, at least, $d_{\mathrm{Raman-HHG}} = (w_0/\lambda_0) [1/\sqrt{\ell}] (1/n)$ within the paraxial approximation, at least a factor of $\sqrt{n}$ smaller than $d_{\mathrm{HHG}}$. The high OAM harmonics could be detected experimentally by using computer generated holograms (as in~\cite{bib:mair_nature_2001}) and will become spatially separated in vacuum due to the differences in their group velocities~\cite{bib:giovannini_science_2015}. The mechanism can be generalised for other complete sets of solutions of the paraxial wave equations, such as the Hermite-Gaussian basis. Because OAM harmonics generation is a result of the wavelike nature of the plasma wave as an optical element, our work may be lead to new types of spiral phase elements capable of enhancing OAM manipulation.

\section{\textbf{Acknowledgments}}
The authors acknowledge fruitful discussions with Dr. Fabien Qu\'er\'e. We acknowledge PRACE fro access to resources on Fermi supercomputer (Italy). Work supported by the European Research Council through the inPairs project and by the EU (EUPRAXIA grant agreement 653782). We acknowledge PRACE for access to resources on Fermi supercomputer (Italy).

\end{document}